\documentclass[submission, Proceedings]{SciPost}
\hypersetup{
    colorlinks,
    linkcolor={red!50!black},
    citecolor={blue!50!black},
    urlcolor={blue!80!black}
}

\usepackage[bitstream-charter]{mathdesign}
\DeclareSymbolFont{usualmathcal}{OMS}{cmsy}{m}{n}
\DeclareSymbolFontAlphabet{\mathcal}{usualmathcal}

\usepackage{siunitx}
\usepackage{subcaption}

\renewcommand{\eqref}[1]{(\ref{#1})}

\newcommand{\figref}[1]{Figure~\ref{#1}}

\numberwithin{equation}{section}
\numberwithin{figure}{section}
\numberwithin{table}{section}

\begin{document}

% For convenience during refereeing (optional),
% you can turn on line numbers by uncommenting the next line:
%\linenumbers

% =========================================================================

\begin{center}{\Large \textbf{
UCN, the ultracold neutron source -
neutrons for particle physics\\
}}\end{center}

% TODO: write the author list here. Use initials + surname format.
% Separate subsequent authors by a comma, omit comma at the end of the list.
% Mark the corresponding author with a superscript *.
\begin{center}
Bernhard Lauss\textsuperscript{1$\star$} and
Bertrand Blau\textsuperscript{1} 
\end{center}

\begin{center}
{\bf 1} Paul Scherrer Institut, CH-5232 Villigen PSI, Switzerland
\\
* bernhard.lauss@psi.ch
\end{center}

%==================================
\begin{center}
\today
\end{center}

\definecolor{palegray}{gray}{0.95}
\begin{center}
\colorbox{palegray}{
  \begin{tabular}{rr}
  \begin{minipage}{0.05\textwidth}
    \includegraphics[width=24mm]{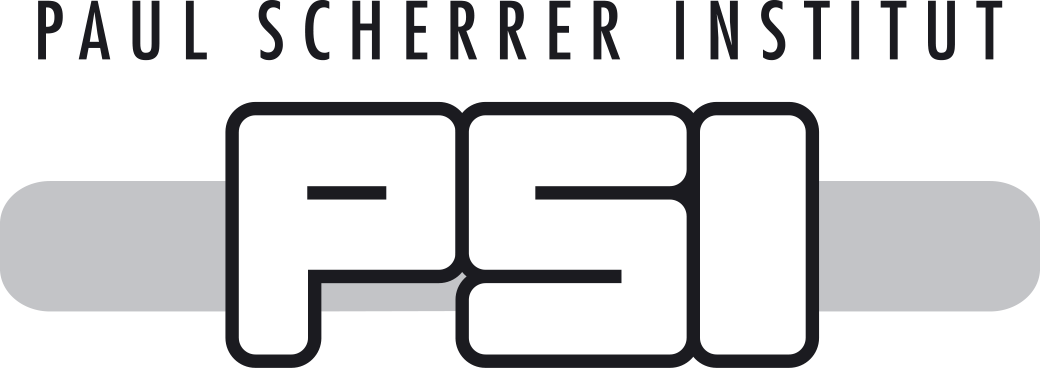}
  \end{minipage}
  &
  \begin{minipage}{0.82\textwidth}
    \begin{center}
    {\it Review of Particle Physics at PSI}\\
    \doi{10.21468/SciPostPhysProc.2}\\
    \end{center}
  \end{minipage}
\end{tabular}
}
\end{center}
%=================================================

\section*{Abstract}
{\bf \boldmath Ultracold neutrons provide a unique tool for the study of neutron
  properties.  An overview is given of the ultracold
  neutron (UCN) source at PSI, which produces the highest UCN intensities
  to fundamental physics experiments by exploiting the high
  intensity proton beam in combination with the high UCN yield in
  solid deuterium at a temperature of 5\,K.  We briefly list important
  fundamental physics results based on measurements with neutrons at
  PSI.  }

\setcounter{section}{4}
\label{sec:UCNsource}

\subsection{Introduction}
\label{UCNsource:intro}

Ultracold neutrons (UCNs) are at the lowest end of the neutron
energy spectrum, 
with kinetic energies below about 300\,neV, corresponding
to velocities below 8\,m/s, and
to temperatures below 4\,mK. Hence they are called ''ultracold``.
This energy is the same as the neutron optical potential of certain
materials. Thus material bottles can be used to store UCNs. 
This energy also corresponds to the potential difference 
of a neutron raised by 3 meters in the Earth's gravitational field, 
and also to the potential difference of a 5 Tesla magnetic field 
gradient acting on the neutron magnetic moment.
Thus, UCNs can be relatively easily confined and manipulated.
Therefore, they are a unique tool
to study the properties of the neutron itself.
The highest UCN intensities are needed to reach the 
highest sensitivity range in fundamental physics 
experiments; the most prominent such experiment
is the search for a permanent electric dipole moment of the 
neutron (nEDM)~\cite{Pendlebury2015,Abel2020}.

% ultracold neutrons

The idea to build an intense UCN source
at PSI 
was formulated in the late 1990's. 
The UCN project was initiated 
and realized under the leadership of Manfred Daum.
The technical design presented in 2000~\cite{UCN1999,UCN2000}
was based on earlier studies in 
Russia~\cite{Pokotilovski1995,Serebrov1997,Serebrov2000}
and a successful operation of a solid-deuterium based UCN source at
the Los Alamos National Laboratory~\cite{LANL2000}.
The main scientific goal was 
to push the sensitivity of the nEDM search to a new level.
%%a new high sensitivity nEDM search.
%
%
Several pioneering experiments by the PSI UCN group  
determining e.g. UCN production 
in solid deuterium~\cite{Atchison2005,Atchison2009b}
and UCN loss cross-sections~\cite{Atchison2005a,Atchison2005b},
paved the way for the final design.
The UCN source was then installed
as the second spallation neutron source
at the PSI high intensity proton facility (HIPA).
After a short test beam period at the end of 2010, 
the UCN source started regular operation in 2011~\cite{Lauss2011,Lauss2012,Lauss2014}
providing UCNs to experiments
at three beam ports.

% a little bit of source history
% and UCN measurements done at SINQ

\subsection{UCN Source Setup}
\label{UCNsource:setup}

% Bild der Quelle

\begin{figure}[htb]
\begin{center}
\includegraphics[width=0.55\textwidth]{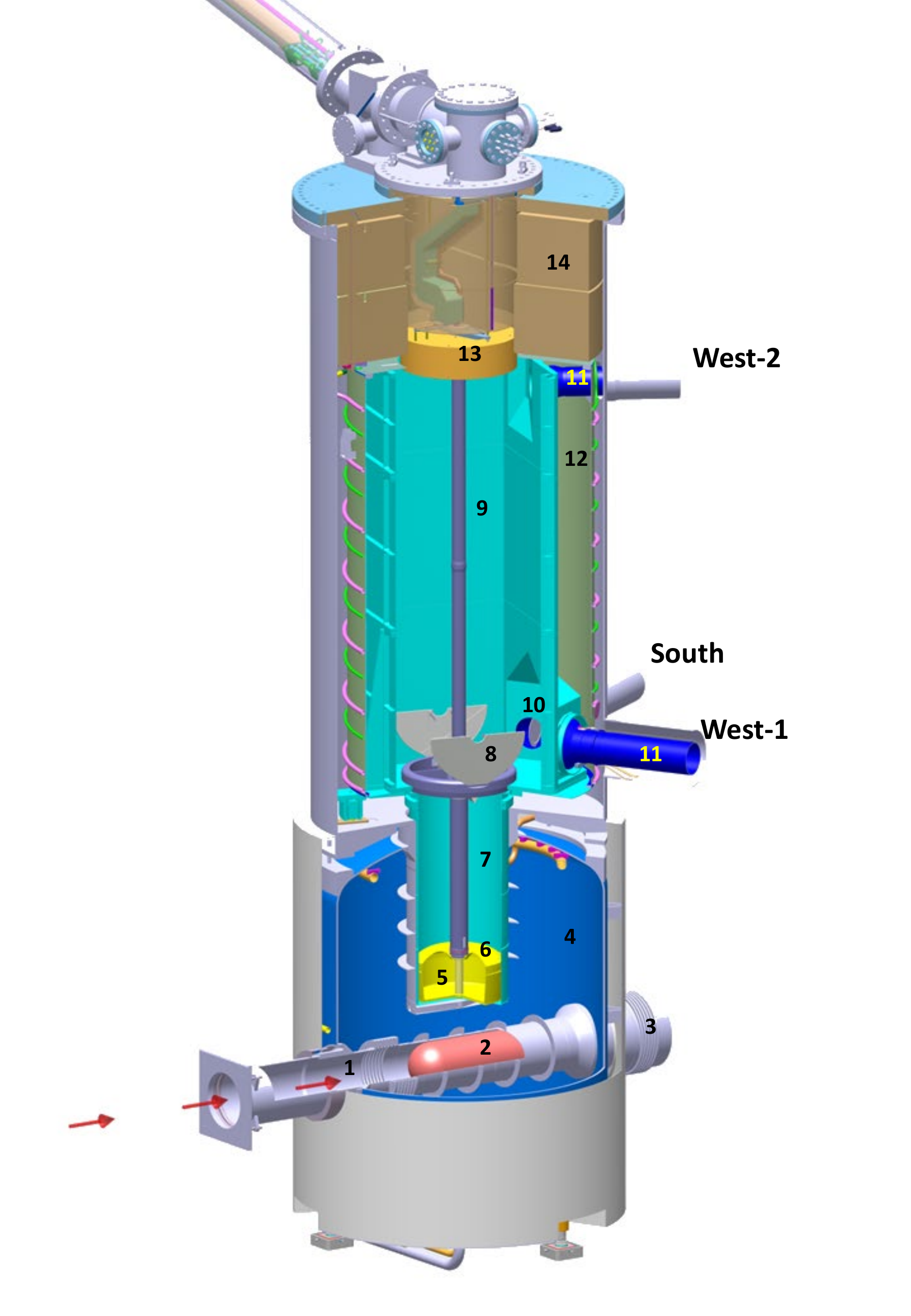}
\caption{
\label{UCNsource:Fig1} 
CAD image of the UCN tank with indicated
parts relevant to UCN production and transport.
1 - proton beam tube,
2 - lead spallation target,
3 - target shielding,
4 - heavy water moderator tank,
5 - D$_2$ moderator vessel, 
6 - lid,
7 - vertical guide,
8 - flapper valve,
9 - storage vessel,
10 - UCN guide shutter,
11 - UCN guide section,
12 - thermal shield,
13 - cryo-pump, at 5\,K,
14 - iron shielding.
%
%Picture from Ref.~\cite{Bison2020} with kind permission of The European %Physical Journal (EPJ).
	}
\end{center}
\end{figure}

\begin{figure}[htb]
\begin{center}
\includegraphics[width=0.8\textwidth]{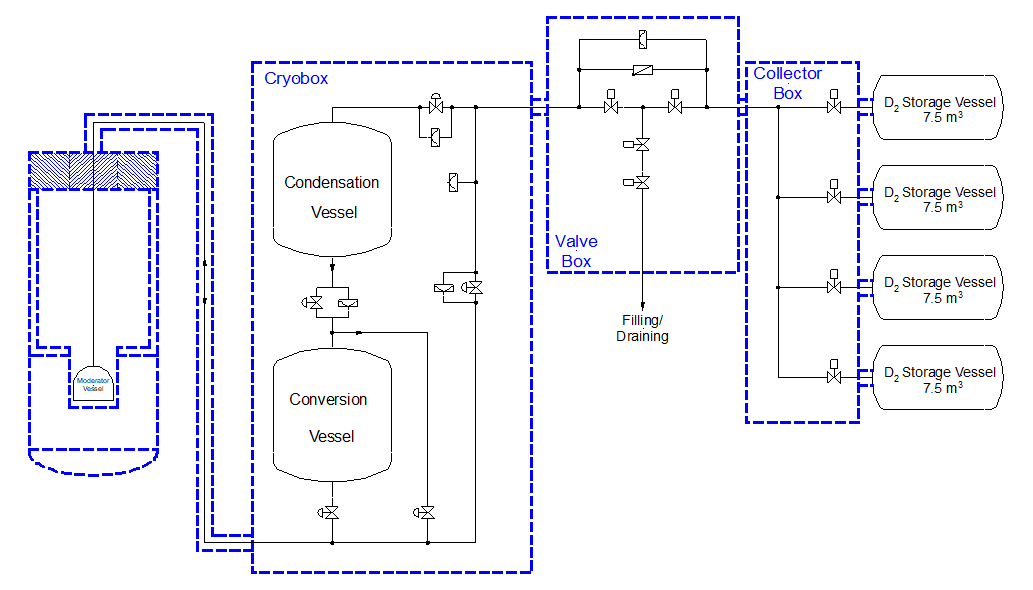}
\caption{
\label{UCNsource:FigD2System} 
Schematic view of the subsystems needed for the
preparation of the solid ortho-deuterium (see text).
}
\end{center}
\end{figure}

The PSI UCN source operates in the following way:
The 590\,MeV, 2.4\,mA proton beam is deflected by a 
kicker magnet~\cite{Anicic2005} for 
up to 8\,s onto the lead spallation target 
(label 2) in \figref{UCNsource:Fig1})~\cite{Wohlmuther2006}.
In a spallation reaction between a lead nucleus and a 590~MeV proton, 
an average of 8 free neutrons is produced~\cite{Becker2015}. 
The neutrons are thermalized in the surrounding 
heavy water (label 4). 
The central moderator vessel (label 5) contains
solid deuterium (sD$_2$) at a temperature of 5\,K, which serves 
as both a cold moderator and as the UCN production medium.
The cryogenics system needed for the manipulation, cooling 
and freezing of the deuterium~\cite{Anghel2008} is 
shown in \figref{UCNsource:FigD2System}.
UCNs exit the moderator vessel through a 
thin aluminum
lid (label 6 of \figref{UCNsource:Fig1}) into a vertical guide where
the energy boost from the sD$_2$ surface~\cite{Altarev2008} 
is lost by gravity. 
The flapper valve (label 8) of the  1.6\,m$^3$ large 
storage vessel is closed at the
end of the proton pulse. 
UCNs trapped in the storage vessel are delivered 
via about 8\,m long neutron guides
(label 11) to three beam ports,
named West-1, South and West-2, with the latter extracting
UCN from the top of the storage vessel.
Great attention was spent on quality checks of all elements,
and extensive tests were performed before installation,
e.g. the cryo-performance of several parts, 
most importantly
the flapper valves (label 8) and UCN guide shutters (label 11). 
The UCN transport performance
of all UCN guides~\cite{Blau2016} 
was confirmed prior to their installation.
The overall neutron optics performance was later analyzed and
understood in terms of a detailed simulation of the entire 
UCN source~\cite{Bison2020}.

\begin{figure}[htb]
\begin{center}
\includegraphics[width=0.45\textwidth]{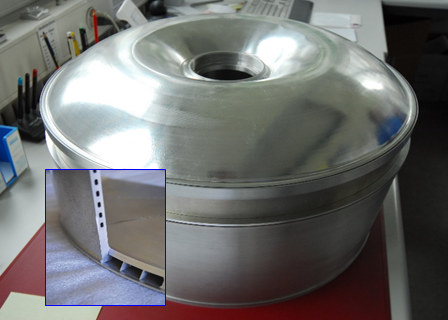}
\includegraphics[width=0.45\textwidth]{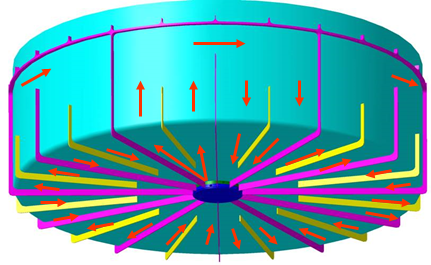}
\caption{
\label{UCNsource:D2vessel} 
a) View of the moderator vessel with a cut insert view
from a test vessel wall.
b) Schematic view of the He coolant flow in the vessel
demonstrating the sectional cooling.
}
\end{center}
\end{figure}

The core of the UCN source is the solid deuterium moderator, 
which also serves as a
UCN converter at a temperature of 5\,K.
The 30 liters of solid D$_2$  
require very careful preparation
in order to achieve optimal UCN output.
A schematic view of the involved subsystem is shown in
\figref{UCNsource:FigD2System}.
Preparation starts from the 
30\,m$^3$ ultra clean
and isotopically pure D$_2$ gas, 
stored in large tanks at ambient temperature, 
which is slowly
transferred by freezing into the 40 liter copper-made 'condensation' vessel. 
The D$_2$ is then slowly liquefied and transferred by gravity into the
'conversion' vessel at about 20\,K where
an ortho-D2 concentration of about 97\% 
is achieved within 24\,h by 
means of a spin-flip process
on Oxisorb$^{\textregistered}$, a chromium-oxide-based catalyzer material.
Raman spectroscopy is used to check the 
ortho-D$_2$ concentration~\cite{Bodek2004,Hild2019}, which
rises up to above 99\% during longer operation periods.
Once the required ortho concentration is reached in the 
conversion vessel, the liquid D$_2$ is transferred 
by gravity through a 10\,m-long cold transfer line 
into the moderator vessel. 
Here it is slowly solidified 
over several days to achieve
a good ice quality and, consequently, 
a high UCN output.
The moderator vessel, shown in \figref{UCNsource:D2vessel}a, 
is entirely made from AlMg3 with special coolant channels
for the supercritical He cooling fluid at 4.7\,K.
These channels enter in the center of the vessel 
and direct the He stream to the outside wall, up and back 
in 8 separated sections,
as schematically depicted in
Fig.~\ref{UCNsource:D2vessel}b.

\begin{figure}[htb]
\begin{center}
\includegraphics[width=0.75\textwidth]{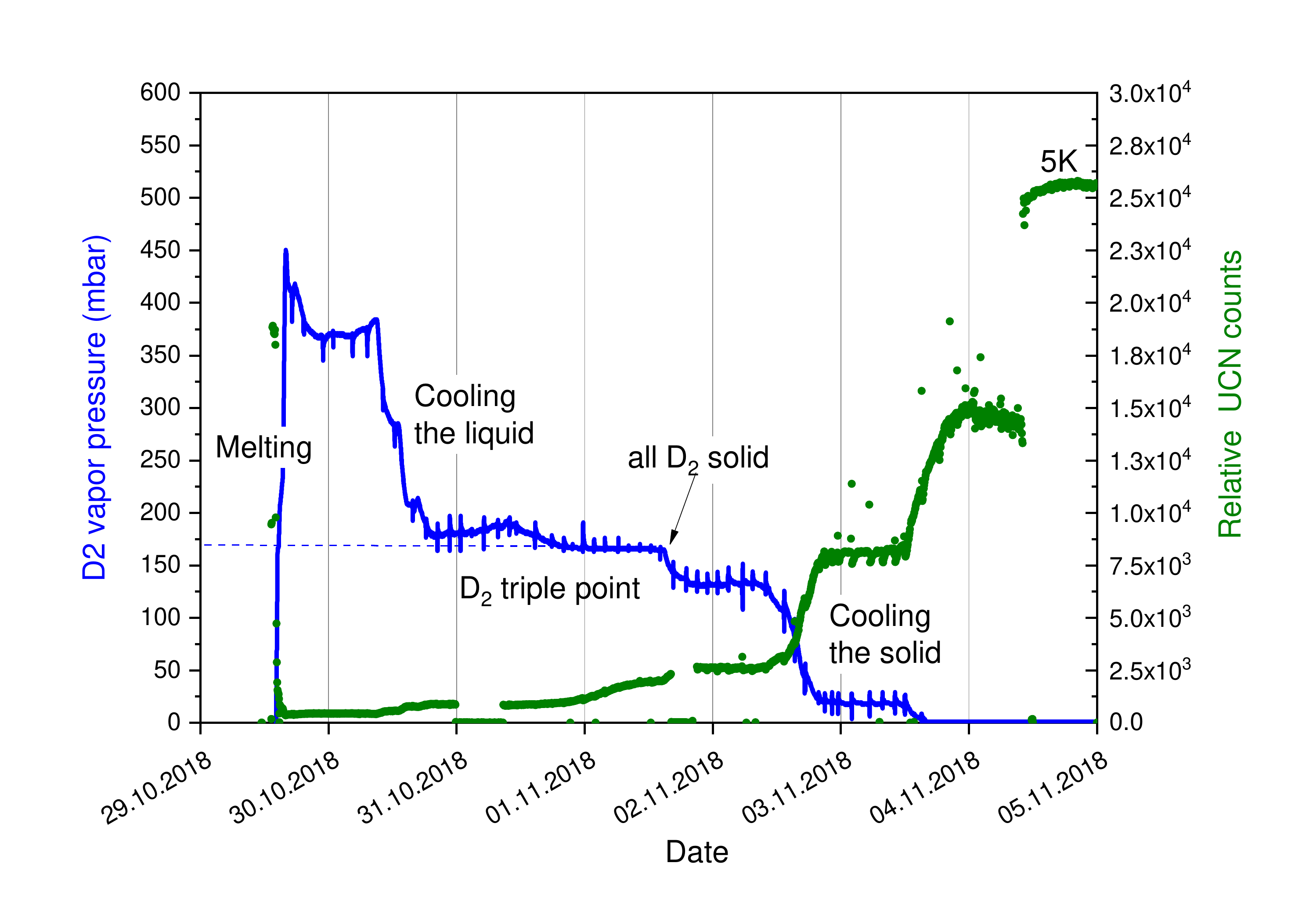}
\caption{
\label{UCNsource:D2-freezing} 
The observed behavior during the slow freezing of the deuterium. 
The vapor pressure of the D$_2$ (blue line) indicates the
D$_2$ temperature.
The D$_2$ was fully melted. 
When it reaches about 400\,mbar vapor pressure, cooling starts
and the D$_2$ slowly approaches the triple point
at 171\,mbar (horizontal dashed line).
Here the D$_2$ solidifies. 
When the solid D$_2$ is further cooled down to 5\,K 
the vapor pressure drops well below 10$^{-2}$\,mbar.
The large increase in UCN output shown by the green bullets
demonstrates the strong reduction in UCN losses within the D$_2$.
%
%The D$_2$ solidifies and the solid is further cooled, 
%indicating a sub 10$^{-2}$\,mbar pressure when reaching 5\,K. 
%
}
\end{center}
\end{figure}

The delivered UCN intensity reflects the quality of 
the achieved solid deuterium, likely a mosaic crystal with many defects
and cracks, 
as was shown in the 
pioneering UCN
experiments~\cite{Atchison2005,Atchison2009b,Atchison2005a,Atchison2005b}. 
Slow freezing is crucial 
in the preparation process of the source.
\figref{UCNsource:D2-freezing} shows the
typical UCN intensity behavior (green line) during such a 
slow freezing process.
The vapor pressure (blue line) 
which is a direct measure of the D$_2$ (surface) temperature
decreases from above 400\,mbar (liquid D$_2$)
to the triple point at about 171\,mbar, where the liquid D$_2$ solidifies.
After solidification the vapor pressure rapidly decreases below 
10$^{-2}$\,mbar.
The UCN output shows the opposite behavior.
UCN loss processes dominate
at higher temperatures, especially in the liquid D$_2$ 
and the high-density vapor located above the D$_2$.
Once 5\,K are reached, thermal losses are minimized and
the UCN output is at its maximum.
%largest, 
%and in addition increasing with growing 
%ortho-D$_2$ concentration and improved crystal quality,
%i.e. less cracks or cavities which can trap UCN.

\subsection{UCN Source Performance}
\label{UCNsource:performance}

\begin{figure}[htb]
\begin{center}
\includegraphics[width=0.47\textwidth]{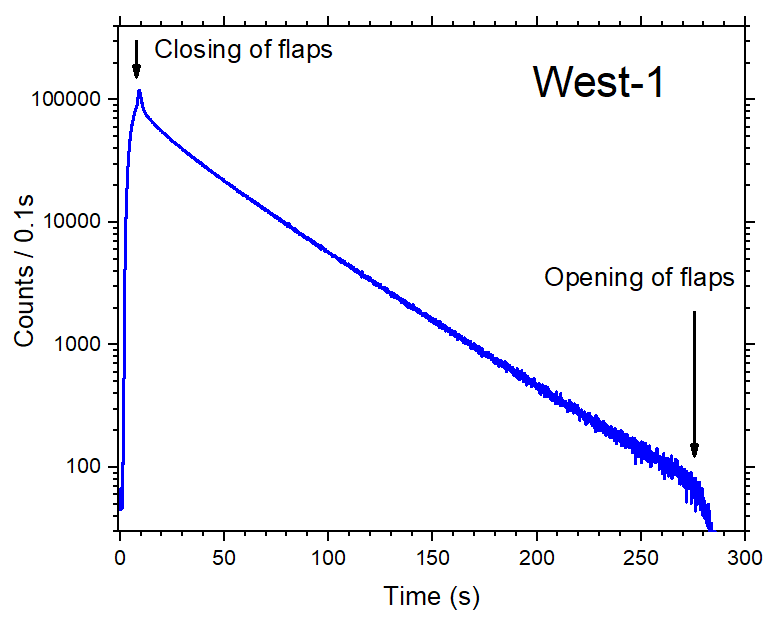}
\includegraphics[width=0.47\textwidth]{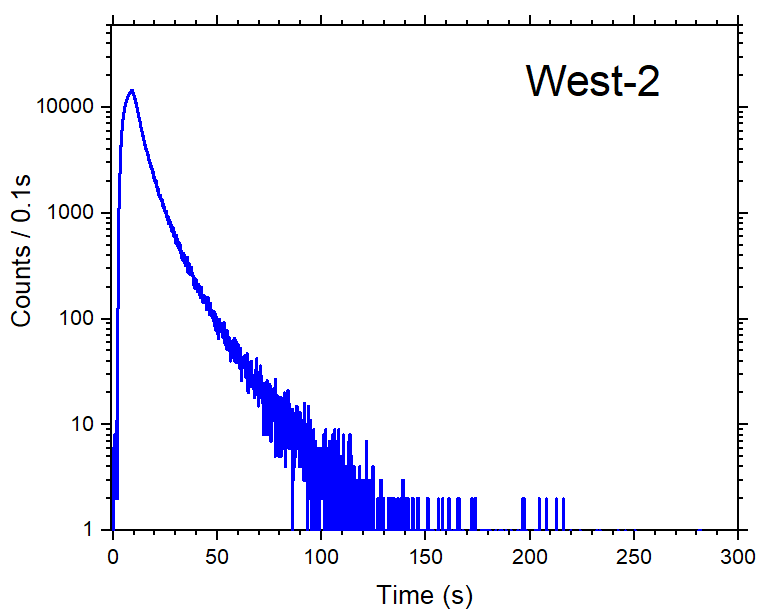}
%%% Rate Decline
\caption{
\label{UCNsource:Fig2} 
a) UCN counts after one proton beam pulse at the West-1 beam port.
Closing and opening of
the flaps refers to the central flapper valves.
b) Same as a) but at the West-2 beam port.
}
\end{center}
\end{figure}

An important performance parameter is the number of UCNs
delivered at a beam port in a given time interval
as this determines the number of UCNs available in an experiment.
The typical time structure of UCNs for a proton 
beam pulse is  shown in \figref{UCNsource:Fig2}.
The flaps of the storage vessel open before the proton beam hits the
spallation target and their closing time is optimized with respect to
the end of the proton pulse to provide the maximum number of UCNs to
the experiments.
The measured exponential decay of the UCN count rate 
at the West-1 beam port, \figref{UCNsource:Fig2}a),
has a time constant of about 30\,s,
reflecting the emptying time of the central storage vessel through 
the West-1 guide into the UCN detector.
The UCN rate at the South beam port behaves identically.  If all shutters
to the UCN guides remain closed on the storage vessel, the storage
time constant for UCNs trapped inside the vessel is about 90\,s.  At
the end of the filling/extraction period, which is typically 300\,s
long, the flaps are re-opened to be ready for the next proton beam
pulse.

\figref{UCNsource:Fig2}b shows the UCN rate 
observed at the West-2 beam port located
230\,cm above the bottom of the storage vessel~\cite{Bison2020}. 
The faster exponential decay demonstrates
that the UCNs with energies high enough to reach 
up to 230\,cm, are quickly 
drained through that port.
%
%The standard operating pulse period is 300\,s.
%
The total number of UCNs delivered at the West-1 or South beam port
was has been up to
45 million at the best operating conditions.
The total number of delivered UCNs
depends on the status of the solid deuterium, and
was increased over the years with 
improvements in the operating conditions.

Several studies to understand all aspects of the UCN source 
have been conducted since its inauguration.
The proton beam current and position is constantly monitored 
online with beam monitors.
Neutron production and thermalization were checked 
using neutron activation measurements on gold.
The observed activation was well reproduced in detailed 
neutron transport simulations using MCNP~\cite{Becker2015}.
Neutron moderation was studied using tritium production in 
the solid D$_2$ moderator~\cite{Hild2019}.
The high ortho D$_2$ concentration and the high isotopic purity
of 0.09\% H atoms (bound in HD molecules)
of the D$_2$ was confirmed~\cite{Hild2019}.

UCN transport from production in the solid deuterium 
to a beam port has been carefully studied as is 
 detailed in the 
thesis works~\cite{Goeltl2012,Ries2016}.
Many geometry details were put into 
a full simulation model and 
the simulation results then matched well with
observations~\cite{Bison2020}.

\begin{figure}[htb]
\begin{center}
\includegraphics[width=0.47\textwidth]{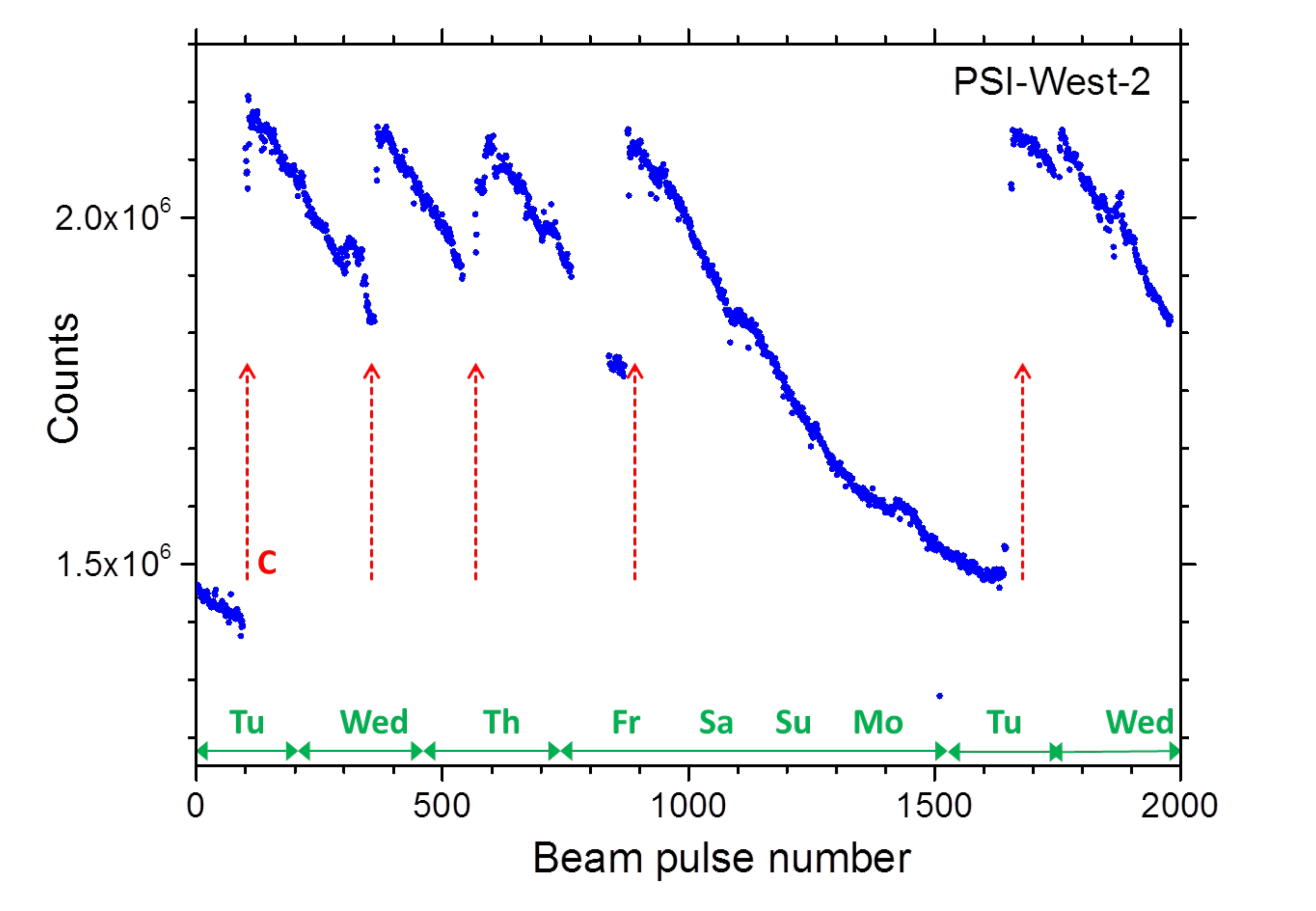}
\includegraphics[width=0.43\textwidth]{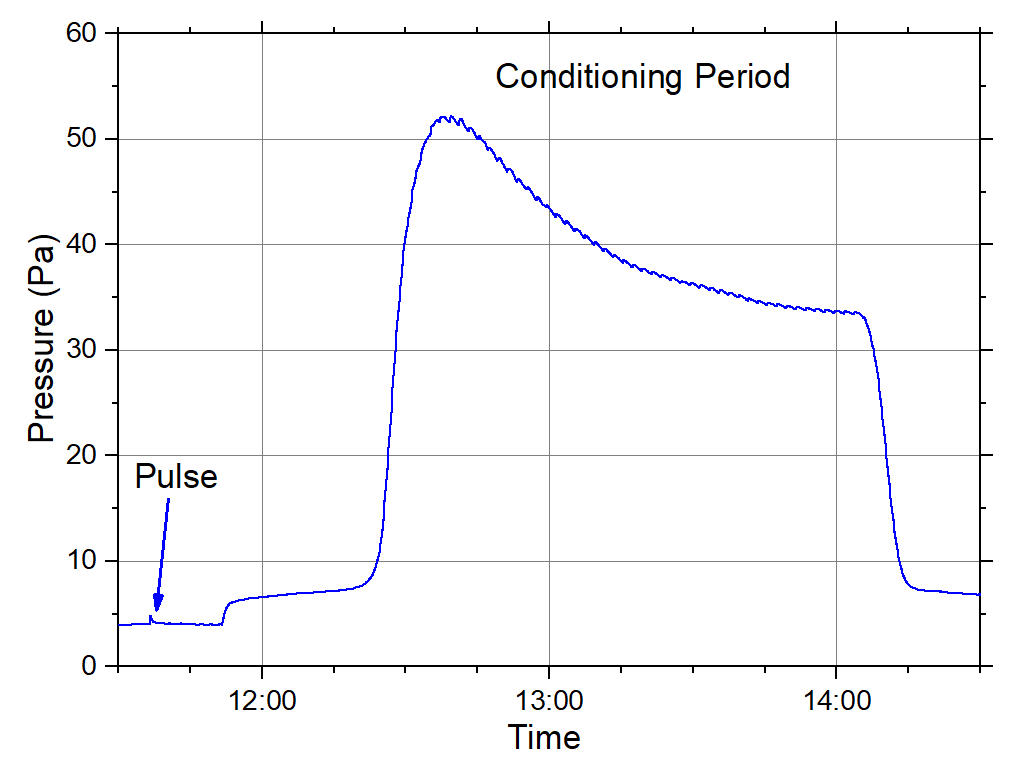}
%
%label{UCNsource:Frost} 
\caption{
a) The UCN count rate behavior as observed over a 9-day
operating period.
The drop is interpreted as a frost effect.
The count rate increases to the original level
when the conditioning procedure is applied
at the times depicted by the dashed arrows.
Figure from~\cite{Bison2020} with kind permission 
of The European Physical Journal (EPJ).
b) Vapor pressure of the solid deuterium surface in the moderator vessel 
during a full conditioning cycle. 
}
\label{UCNsource:Frost} 
\end{center}
\end{figure}

The measured integral
UCN intensity per beam pulse also shows a time dependence
on the scale of several hours to days, 
which considerably decreases the average UCN output.
After several studies, a temperature-cycling 
procedure, called ``conditioning'', was developed 
that gets rid of the accumulated losses 
and 
regains maximum UCN intensity.
This UCN count rate behavior is shown in 
\figref{UCNsource:Frost}a, 
where the times when the conditioning procedure was
applied are labeled by the vertical arrows.
\figref{UCNsource:Frost}b shows the measured deuterium vapor
pressure in the moderator vessel during a 2-hour conditioning process. The rise in vapor pressure during a proton
beam pulse, noted with the blue arrow, is minuscule.  The rise
during temperature cycling is up to about 50\,Pa, depending on the
total operation time since the previous conditioning.  This is far
below the triple-point pressure of 171\,mbar and is due to
sublimation, movement and resublimation of surface molecules during
conditioning.  Interesting enough, full rate recovery occurs.

\begin{figure}[htb]
\begin{center}
\includegraphics[width=0.55\textwidth]{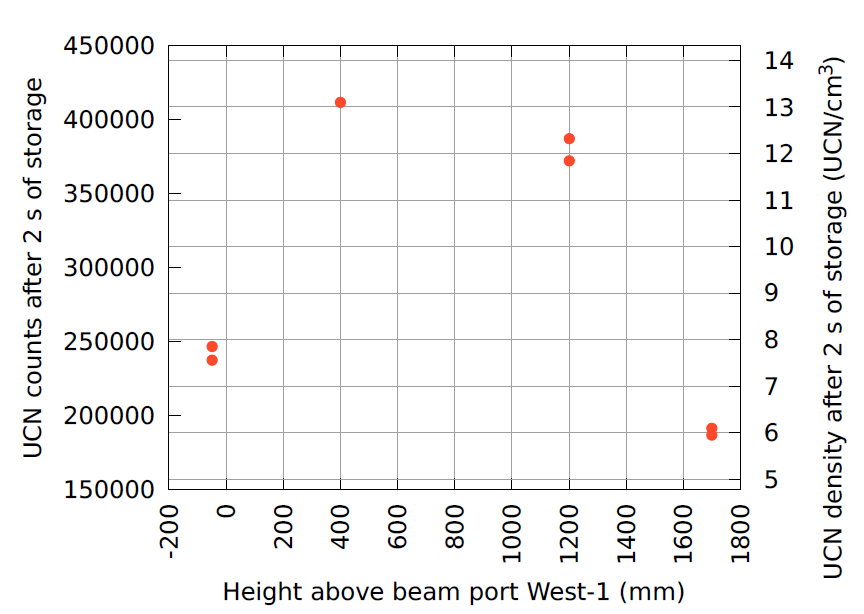}
\caption{
\label{UCNsource:density} 
The UCN density measured at different heights with respect
to the West-1 beam port.
Picture from~\cite{Ries2016}.
	}
\end{center}
\end{figure}

One of the key characteristics of a
UCN source is the UCN density that can be achieved in 
a given storage vessel.
A stainless steel 'standard UCN storage vessel'
with a volume of 20\,liters~\cite{Bison2016} was built.
This bottle was used to characterize the 
height-dependent 
UCN density at the West-1 beam port~\cite{Ries2016}.
The UCN density peaks around 50\,cm above the beam port as
shown in \figref{UCNsource:density}.
This standard bottle was then used to characterize UCN densities of
other sources in a comparable way~\cite{Ries2016,Bison2017,Kahlenberg2017}.  As
a result it has been shown that the PSI source provides world-leading
performance to UCN storage experiments.

\begin{figure}[htb]
\begin{center}
\includegraphics[width=0.5\textwidth]{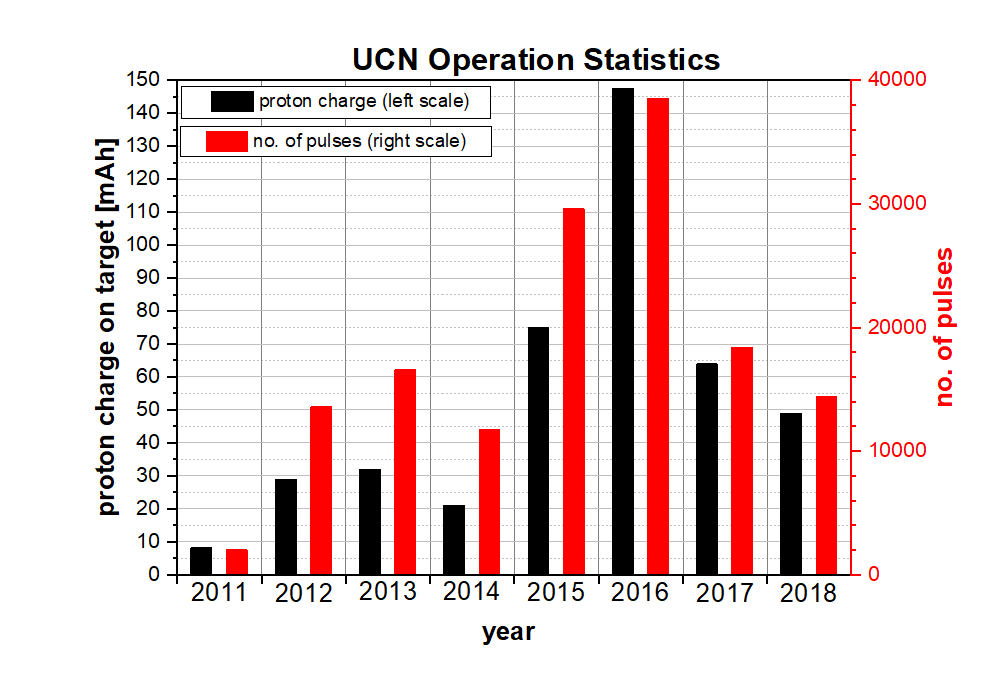}
\caption{
\label{UCNsource:statistics} 
Annual statistics of the first operating years of the UCN
source showing total accumulated beam current on target
(black bars) and
number of beam pulses (red bars)
on the UCN spallation target.
}
\end{center}
\end{figure}

The PSI UCN source has been operating since 2011 on a regular
schedule, mainly providing UCNs to the nEDM experiment.  The yearly
operation can be characterized by the integral of the proton beam current onto
the UCN spallation target and the number of proton beam pulses, shown in \figref{UCNsource:statistics}.  The peak in
2016 was driven by the main data taking period of the nEDM experiment.
The lower numbers in the subsequent years are due to longer periods of
solid deuterium studies for UCN source improvements, which needed
longer times with fewer proton beam pulses for performance checks.

%%%%%%%%%%%%%%%%%%%%%%%%%%%%%%%%%%%%%%%%%%%%%%%%%%%%%%%%%%%%%%

\subsection{Physics results at the UCN source}
\label{UCNsource:results}

The  construction of the UCN source at PSI was driven by
the experiment to search for a neutron electric dipole moment.
The resulting new nEDM limit was published in 2020~\cite{Abel2020}.
Significant physics results were also obtained
on neutron properties and effects:
%%%%
%%%%
\begin{itemize} 
\setlength\itemsep{-1.0em}
\item a precision measurement of the mercury-to-neutron 
magnetic moment ratio~\cite{Afach2014};\newline
\setlength\itemsep{-1.0em}
%\item observation of gravitationally-induced vertical striation of polarized ultracold neutrons by spin-echo spectroscopy~\cite{Afach2015PRL}
\item spin-echo spectroscopy with ultracold neutrons~\cite{Afach2015PRL}
\setlength\itemsep{0.0em}
\item measurement of gravitational depolarization of ultracold neutrons~\cite{Afach2015PRD}
\end{itemize}
%%%%%%%%
%%%%%%%%%%
%%%%%%%%%%%
and on physics beyond the Standard Model:
\begin{itemize}
\setlength\itemsep{-1.0em}
\item a limit for spin-dependent forces mediated by axion-like 
particles~\cite{Afach2015Exotic};\newline
\item the first laboratory
limit for oscillating electric dipole 
moments~\cite{Abel2017};\newline
\item new limits for mirror-neutron oscillations in mirror 
magnetic fields~\cite{Abel2021}.
\end{itemize}
Some of these results are treated in Section~18~\cite{section18} and
Section~19~\cite{section19} of this volume.

\subsection{Particle physics at the SINQ}
\label{UCNsource:SINQ}

The UCN source was conceived and built for research
in fundamental neutron physics. 
However, the first spallation neutron source built at PSI was
the SINQ facility~\cite{SINQ}.
% {\bf \color{red}(cite ??)}.
%
While mainly dedicated to neutron scattering instruments,
it has also been used as
a polarized cold-neutron beam line 
for fundamental neutron physics.
The 'FUNSPIN' beam line~\cite{Bodek2000} (now called 'BOA')
provided 6$\times$10$^8$ neutrons 
cm$^{-2}$s$^{-1}$mA$^{-1}$
with 95\% polarization~\cite{Zejma2005}.

The main physics results came from a series of measurements by the nTRV collaboration of neutron decay parameters.
A precise determination of electron-neutron correlation coefficients
{\it R} and {\it N}
provided a precise test of the Standard Model and
a search for exotic scalar and tensor interactions in neutron 
decay~\cite{BanBialek2009,Kozela2009,Bodek2011b,nTRV2012}.

Another experiment produced a new measurement
of the spin-dependent doublet neutron-deuteron 
scattering length~\cite{Piegsa2007,Piegsa2009}.
A Ramsey-type experiment resulted in an
upper limit on the strength of an axial coupling constant 
for a new light spin 1 boson in the millimeter range~\cite{Piegsa2012}.

%  UCN source preparation experiments
Finally, we note the importance of the
FUNSPIN beamline for many measurements 
conducted in preparation of the UCN 
source
where many parameters of UCN production and
loss were determined~\cite{Atchison2005,Atchison2009b,Atchison2005a,Atchison2005b,Atchison2007,Atchison2007a,Atchison2011}.

%%%%%%%%%%%%%%%%%%%%%%%%%%%%%%%%%%%%%%%%%%%

\subsection{Summary}
\label{UCNsource:summary}

A high-intensity source for ultracold neutrons,
designed and built at PSI, has been operating since 2011.
The layout, operation and performance are described.
Some observations on the solid deuterium
converter and its surface conditions are presented. 
Finally, a list of physics results in 
fundamental neutron physics results achieved with the UCN source 
and SINQ is given.

%%%%%%%%%%%%%%%%%%%%%%%%%%%%%%%%%%%%%%%%%%%%%%%%

\subsubsection*{Acknowledgments}

Building the UCN source at PSI required the dedicated
long-term support by many individuals and support groups at PSI that
worked for several years together within the UCN source team.
We especially acknowledge the invaluable contributions of all former
and present members of the BSQ group and the UCN physics group.

\bibliography{ucn}

\nolinenumbers

\end{document}